\documentclass[%
 aip,
 amsmath,amssymb,
 reprint,%
]{revtex4-1}

\usepackage{graphicx}
\usepackage{dcolumn}
\usepackage{bm}

\usepackage{graphicx}
\usepackage{amsmath}
\usepackage{amsfonts}
\usepackage{latexsym}
\usepackage{amsbsy}
\usepackage{epstopdf}
\usepackage{bm}%
\usepackage{subfig}
\usepackage[utf8]{inputenc}
\usepackage[T1]{fontenc}
\usepackage{mathptmx}
\usepackage{etoolbox}

\makeatletter
\def\@email#1#2{%
 \endgroup
 \patchcmd{\titleblock@produce}
  {\frontmatter@RRAPformat}
  {\frontmatter@RRAPformat{\produce@RRAP{*#1\href{mailto:#2}{#2}}}\frontmatter@RRAPformat}
  {}{}
}%
\makeatother
\begin{document}

\preprint{AIP/123-QED}
\title{Enhancement of electromagnetically induced transparency based Rydberg-atom electrometry through population repumping}

\author{Nikunjkumar Prajapati}
\affiliation{{National Institute of Standards and Technology, Boulder,~CO~80305, USA}}
\affiliation{Depart. of Phys., University of Colorado, Boulder,~CO~80305,~USA}

\author{Amy K. Robinson}
\affiliation{Depart. of Electr. Engin., University of Colorado, Boulder,~CO~80305,~USA}

\author{Samuel Berweger}
\author{Matthew T. Simons}
\author{Alexandra B. Artusio-Glimpse}
\author{Christopher L. Holloway}
\affiliation{{National Institute of Standards and Technology, Boulder,~CO~80305, USA}}

\date{\today}




\begin{abstract}
We demonstrate improved sensitivity of  Rydberg electrometry based on electromagnetically induced transparency (EIT) with a ground state repumping laser. Though there are many factors that limit the sensitivity of radio frequency field measurements, we show that repumping can enhance the interaction strength while avoiding additional Doppler or power broadening. Through this method, we nearly double the EIT amplitude without an increase in the width of the peak. A similar increase in amplitude without the repumping field is not possible through simple optimization.We also establish that one of the key limits to detection is the photon shot noise of the probe laser. We show an improvement on the sensitivity of the device by a factor of nearly 2 in the presence of the repump field. 
\end{abstract}

\maketitle


Rydberg atom based electrometry is a field that has gained considerable attention in the past decade~\cite{first_ryd_elet_shaff, gor1, holl1, holl2, Fan_2015}. This is primarily due to the large polarizabilty offered by highly excited atoms~\cite{gallagher_book} and their ability to act as a calibrated standard for various measurements~\cite{holl1, si_trace,pow_stand}. For this reason, Rydberg states are used for greatly enhanced field sensing in communications~\cite{Song:19,9054945,8778739,dual_species,waveguide_SA,rydberg_array,comm_sense_fancher,AM_FM_anderson}, power measurements~\cite{pow_stand}, temperature sensing~\cite{Norrgard_2021,Fan_2015,9054945}, and imaging~\cite{terehertz_imaging}. In addition to the enhanced sensing offered by Rydberg atoms, the electric field measurements are traceable to the international system of standards (SI)~\cite{SI_standard} through Planks constant~\cite{gor1, holl1, pow_stand}. 


\begin{figure}[h!]
    \centering
\scalebox{.22}{\includegraphics*{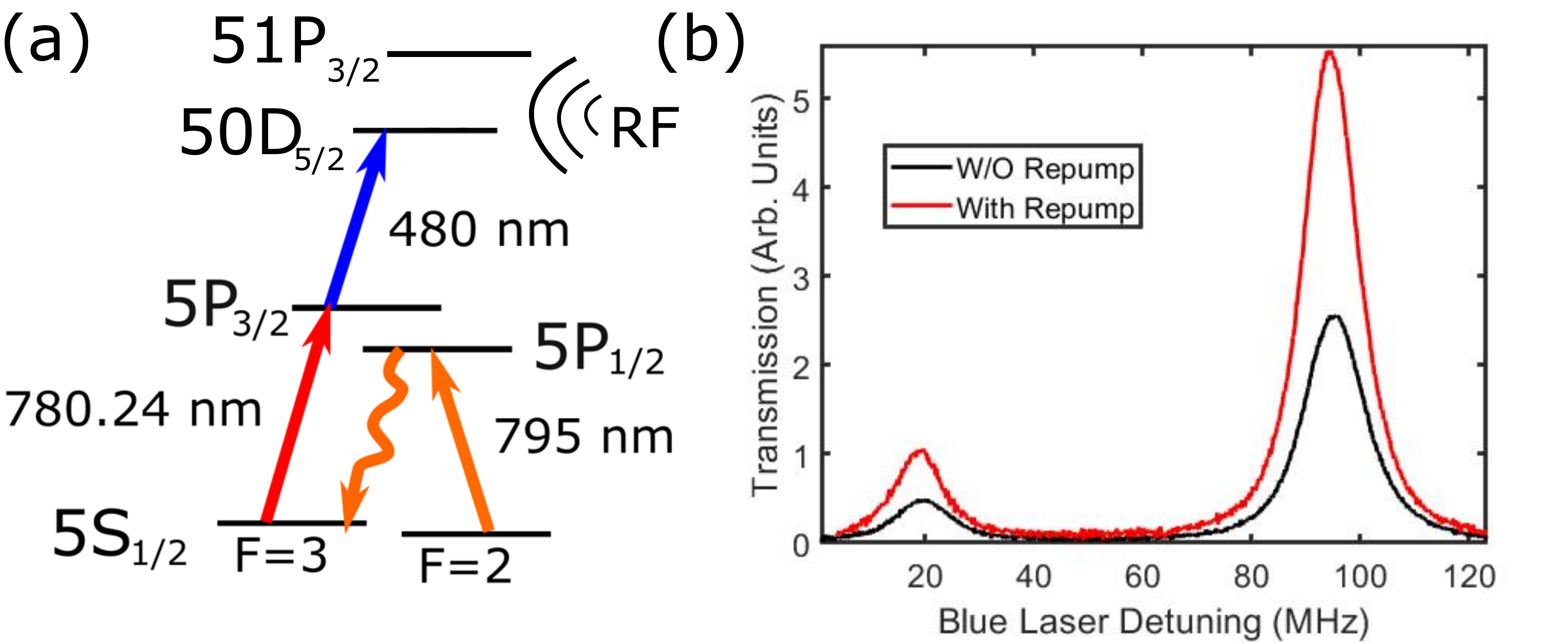}}
\caption{(a) Level diagram of EIT coupling 5S state to the 50D Rydberg state through 5P$_{3/2}$ intermediate state: repumping from unused 5S$_{1/2}$,F=2 state to the 5P$_{1/2}$,F=3 through 5P$_{1/2}$ state in $^{85}$RB.
(b) 780 laser transmission plotted against the coupling laser detuning.}
    \label{fig:repump_effect_sample}
\end{figure}

Many Rydberg electrometry techniques rely on electromagnetically induced transparency (EIT) to probe the interaction between the atoms and the fields of interest ranging from DC-THz. EIT is a nonlinear process that couples two states that have a dipole-forbidden transition with a two-photon resonance utilizing an intermediate state, as shown in Fig.~\ref{fig:repump_effect_sample} (a). EIT results in a narrow resonance feature that is used to probe highly excited states with increased accuracy and precision. With this narrow feature, we can probe the energy level shifts of Rydberg states due to various field atom interactions like the DC Stark shift~\cite{ma2021measurement,plate_cell_sandia}, AC Stark shift~\cite{r12, r13, davecon,PhysRevLett.98.203005}, and Autler-Townes (AT) splitting~\cite{first_ryd_elet_shaff, gor1, holl1, pow_stand,si_trace}, thereby characterizing the field of interest. While this EIT resonance is narrow, it is weak and Doppler limited. Experimental techniques have yet to achieve the expected natural decay width~\cite{si_trace,freq_det,dual_species,pow_stand,9054945,Fan_2015}, ultimately limiting the sensitivity of these Rydberg-based techniques. However, through the use of the AT splitting and a RF local oscillator field, the frequency splitting measurement can be converted into a an modulated amplitude measurement. With a lock-in amplifier, this method results in three orders of magnitude improvement and a field sensitivity of 5.5 $\mu$Vm$^{-1}$Hz$^{-1/2}$ ~\cite{gor3,Jing2020}.


There are several aspects which limit us to this sensitivity in the atom-based mixer approach. The primary effect comes down to the doppler effect and the residual doppler shifts due to the frequency mismatch of the probe (780 nm) and coupling (480 nm) lasers. Because of this, some of the key parameters used to tune the sensitivity of Rydberg sensors are coupled. In order to enhance the amplitude of the EIT peak, the power of the lasers must be increased. However this lends itself to power broadening that ultimately limits the fundamental sensitivity. Alternatively, we can increase the number of atoms in the interaction by increasing the temperature or optical power. However, this increases the doppler broadening and increases the residual doppler mismatch of the probe and coupling fields. In addition to this, the competing effects of increased number of atoms and their collision with the Rydberg atoms cause decoherence and can weaken and broaden the EIT line further~\cite{Fan_2015}.

In this manuscript, we use a ground state repumping to increase the number of Rydberg atoms in the EIT interaction without doppler-, power-, or collisional-broadening from other methods.
Such a method has been used to increase the frequency conversion efficiency for optical beams~\cite{Prajapati:18,Akulshin_2011}, showing an over a factor of 5 increase in conversion efficiency. With a ground state repumping laser, shown by orange ray in Fig.~\ref{fig:repump_effect_sample} (a), we excited atoms from the 5S$_{1/2}$, F=2 state to the 5P$_{1/2}$ state which then decay to the 5S$_{1/2}$, F=3 state. This effectively doubles the number of atoms in the EIT interaction without changing the velocity group of the atoms. This is possible since the rubidium atoms in the interaction region are naturally split between the two hyperfine ground states (5S$_{1/2}$, F=2 and F=3) of the alkali atom. With the repump we are simply incorporating these atoms into the interaction.
Additionally, when optical fields are used to control the populations of the excited states, decays to the non-interacting ground state become more common. These atoms are then trapped in this state due to the dipole selection rules and are naturally reintroduced very slowly through collisional effects. However, by reintroducing these atoms into the interaction with a repump laser, we can add atoms into the interaction and increase the strength of the interaction, shown by red trace in Fig.~\ref{fig:repump_effect_sample} (b). Additionally, these "new" atoms exist in the same doppler profile and do not contribute to a broadening of the EIT width.


In this demonstration, we use Rydberg atoms to measure the field strengths of weak radio frequency (RF) fields. When such an alternating field is resonant with atomic transitions, the energy levels experience a splitting proportional to the field strength and the dipole moment of the transition. The AT splitting becomes more sensitive for higher lying Rydberg states, but is limited for a fixed state by the amplitude and width of the EIT peak. These factors are determined by the power of the lasers, density of the atoms, and the doppler mismatch of the lasers being used. By using a repump field, we can enhance the EIT amplitude line beyond what is possible by only optimizing these parameters.

\begin{figure}[h!]
    \centering
    \scalebox{.22}{\includegraphics*{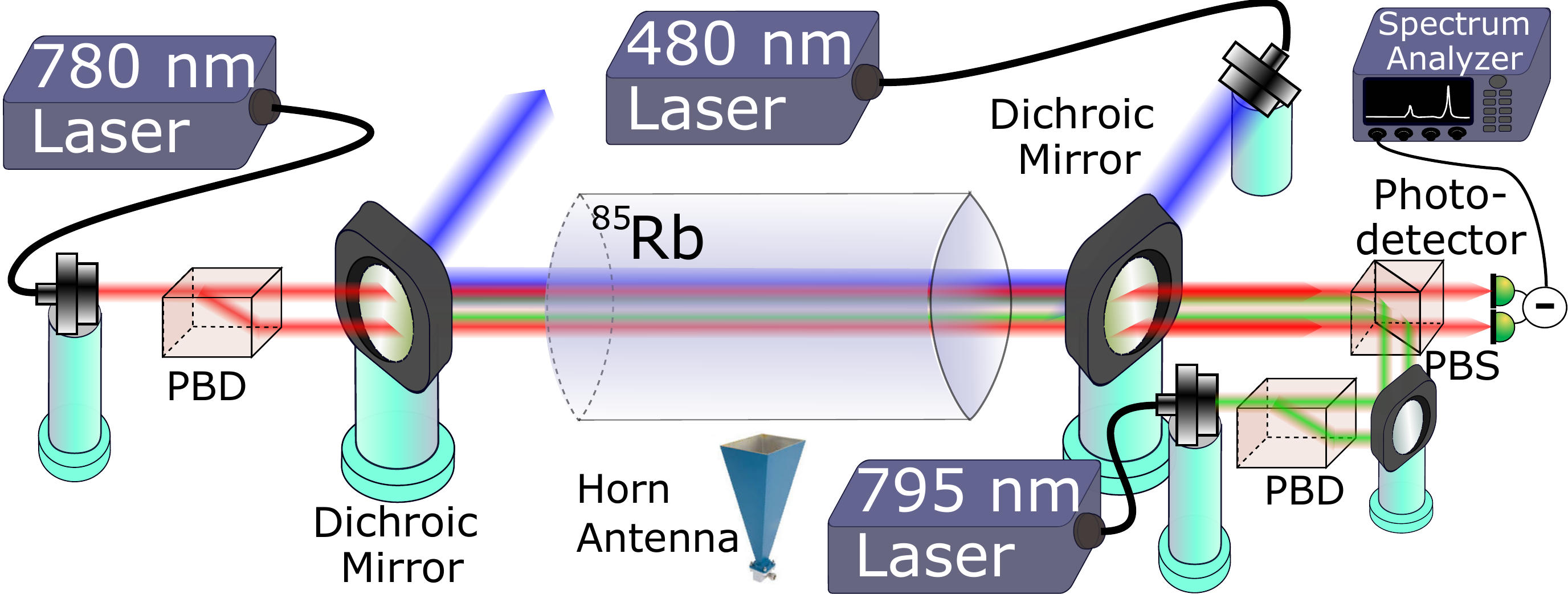}}
    \caption{
    Experimental setup: the 795~nm laser is  used to 
    pump atoms into the hyperfine state of interest. 
The acronyms are polarizing beam displacer (PBD) and polarizing beam  splitter (PBS).}

    \label{fig:setup}
\end{figure}

To observe the effects of the repump field, we look at the alterations to the EIT line due to the repump laser. The experimental apparatus is shown in Fig.~\ref{fig:setup}.
We generate EIT with a 780 nm external cavity diode laser (ECDL) as our probe and a 480 nm laser derived from second harmonic generation of a 960 nm ECDL as our coupling beam. Both lasers are collimated and pass through the rubidium vapor cell in counter-propagating directions to account for the Doppler shift in the ladder configuration. The 795 nm repump optical field is also generated by an ECDL and combined to be co-propagating with the blue coupling laser. While the 795 nm optical field does not introduce wave vector-dependent Doppler effects, we want to avoid the light leakage into the photo-detector (PD) that would add noise. To assure that laser noise is limited, we use balanced detection to cancel common mode noise (discussed further later). The beam widths for the probe and coupling optical fields at the location of the cell are $\approx$2~mm. The large beams were required to counter the power broadening effects while maximizing the number of atoms in the interaction. This was done to maximize the EIT amplitude while maintaining a small width. While even larger beams might show improved results, we were limited by the laser power available in the coupling laser. This limitation arises from the small dipole transition element for the energy levels connected by the coupling laser, thereby increasing the required laser power to achieve the same Rabi frequency. The repump field was diverging entering the cell with a 2.5 $\mu$m width. 

For this demonstration, we tune the 780 nm probe laser to the $5S_{1/2},F=3\rightarrow5P_{3/2},F'=4$ transition and the 480 nm coulping laser to the $5P_{3/2},F'=4\rightarrow50D_{5/2}$ transition. Since this is a two-photon interaction, the coupling laser scan will probe the Rydberg states of interest. When resonant with a Rydberg state, the probe will experience a transmission to readout the Rydberg state and any effects on it, as shown in Fig.~\ref{fig:repump_effect_sample} (b). Additionally, we calibrate our coupling laser scan by taking advantage of the hyperfine structure present the traces in Fig.~\ref{fig:repump_effect_sample} (b). This also allows us to accurately determine the width of the EIT peak and the separation of the peak when measuring the AT splitting. Therefore, we obtain the EIT amplitude and width through a fit of the traces in Fig.~\ref{fig:repump_effect_sample} to gauge the effects of the repump.



The ultimate sensitivity is limited by how well we can determine the AT splitting due to the RF field. The factors determining sensitivity are the EIT amplitude, EIT width, and noise on the measurement. Common methods to increasing EIT signal strength include tuning the beam size, powers, and atomic density, but these have their limitations. For instance, an increase in the probe or coupling optical power will saturate and show only modest improvement. 

\begin{figure}[h!]
    \centering
    \scalebox{.28}{\includegraphics*{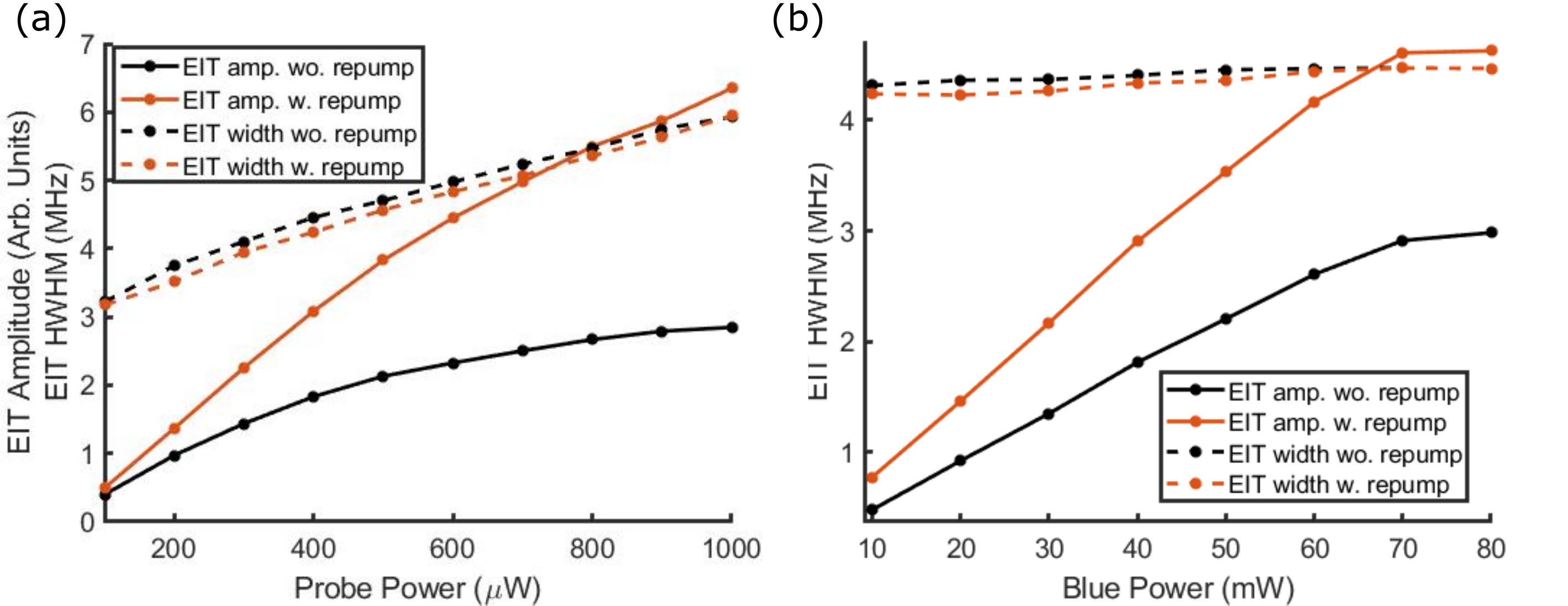}}
    \caption{a) EIT amplitude (solid) and EIT half-width-half-max (HWHM) (dashed) for different probe powers with (red) and without (black) the repump field, coupling fixed to 40 mW. (b) Same as (a), but plotted against coupling field power with probe power fixed to 400 $\mu$W.}
    \label{fig:repump_probe_dependance}
\end{figure}

Figs.~\ref{fig:repump_probe_dependance} (a) and (b) show the effects of the probe power and coupling power, respectively on the EIT amplitude (solid) and widths (dashed) for cases where the repump is present (red) and is not present (black). It can be seen that increasing the probe field power has immediate power broadening accompanied with the amplitude increase, shown by dashed black trace in Fig.~\ref{fig:repump_probe_dependance} (a). This broadening is detrimental when trying to identify the AT splitting. Here, the costs outweigh the benefits. However, with the introduction of the repump field, we can immediately see the EIT amplitude and saturation level increase, shown by solid red trace in Fig.~\ref{fig:repump_probe_dependance}. At 300 $\mathrm{\mu}$W of probe power, we nearly a double the EIT amplitude due to the repump field, but the width of the peak does not change much, shown by dashed traces in Fig.~\ref{fig:repump_probe_dependance}(a). In addition to this, no increase in probe power will reach the same EIT amplitude attained by introducing the repump field due to saturation, solid traces in Fig.~\ref{fig:repump_probe_dependance}(a). 

While the effects of the repump scale nonlinearly with the probe power, the effects are linear with the coupling field power, solid traces in Fig.~\ref{fig:repump_probe_dependance}(b). Also, unlike the case of the probe power, the saturation occurs at the same coupling field power.  This effect is likely due to limited population change from the coupling field, whereas the probe field has direct consequences on the hyperfine split ground states. The effects of the repump field on the EIT amplitude provide over a factor of 2 increase in the amplitude while showing no increase in the width of the EIT resonance for any given probe and coupling field powers. Additionally, repumping requires only a modest amount of power. Even though the repump beam has a larger width than the probe and coupling beams, it saturates the pumping transition near the same power as the probe.
For the remainder of this manuscript, we use a probe field optical power of 550$~\mu\mathrm{W}$, a coupling power of 72$~\mathrm{mW}$, and a repump power of 10$~\mathrm{mW}$ measured before entering the cell. These powers offer roughly a factor of 1.5 increase on the EIT amplitude while not substantially increasing the peak width. While the large repump power is not necessary, the strong saturation helps avoid repump laser intensity fluctuations from being transferred to the probe.

\begin{figure}[h!]
    \centering
        \scalebox{.28}{\includegraphics*{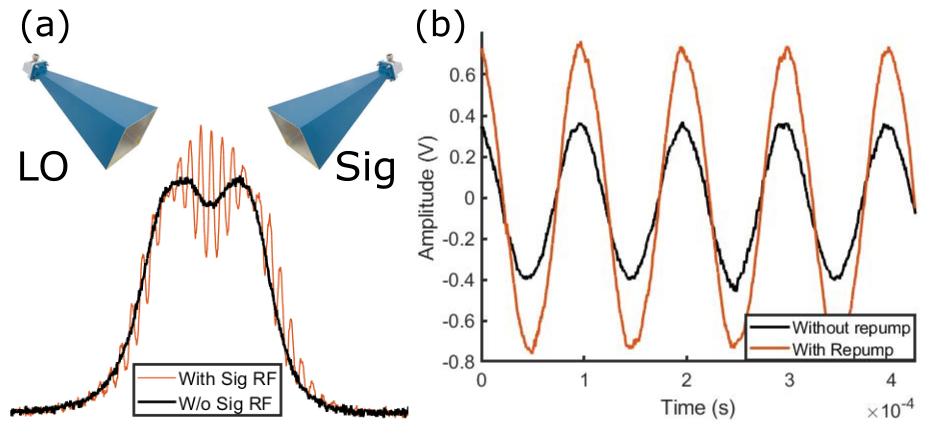}}
    \caption{(a) AT splitting of EIT by LO horn (black trace) and AT splitting amplitude modulated by Sig and LO field (red trace)
    (b) EIT signal with probe and coupling laser locked to resonance showing beat-note from Sig and LO RF fields separated by 10 kHz with (red) and without (black) repump.}
    
    \label{fig:beat_sample}
\end{figure}

The most sensitive RF field detection techniques rely on the concept of the atom-based field mixer~\cite{doi:10.1063/1.5088821,8878963,gor3, Jing2020}. An RF field (say a local oscillator--LO) causes AT splitting, shown by black trace in Fig.~\ref{fig:beat_sample}(a). Then a second RF field interferes with the LO resulting in interference, shown by red trace in Fig.~\ref{fig:beat_sample}(a). While the fields sum in free space, atoms only perceive the field amplitude, thereby mixing the signals and observing the beat-note as a modulation on the AT splitting. For an optimum LO field, the modulation of the AT splitting translates to an amplitude modulation of the EIT peak, shown by Figs.~\ref{fig:beat_sample}(a) and (b). By locking to the EIT peak and measuring the amplitude of this beat-note through a lock-in amplifier, weak signal fields can be detected. Field sensitivity down to $5\mathrm{\mu V(m\sqrt{Hz})^{-1}}$ has been shown with this technique\cite{gor3,Jing2020}.

By introducing a repump field, we can effectively double the strength of the beat-note, shown by red trace in Fig.~\ref{fig:beat_sample} (b). Ideally, this enhancement should increase our sensitivity. We use a lock-in amplifier to detect the strength of the beat-note signal. For this demonstration, we used a LO resonant with the 50$D_{5/2}\rightarrow51P_{3/2}$ Rydberg states that corresponds to a frequency of 17.04340 GHz. The LO was generated using a signal generator (SG) set to an output RF power of -14 dBm. 

The AT splitting provides an immutable  measurement of the E-field traceable to the International System of standards (SI) through Plank's constant through
\begin{equation}
    |E| = \frac{h\cdot\Delta_{AT}}{\wp},
\label{eq:at_effect}
\end{equation}
where $\Delta_{AT}$ is the AT splitting, $\wp$ is the transition dipole moment, $h$ is Plank's constant, and E is the electric field strength.  Using Eq.~(\ref{eq:at_effect}), we determined that this corresponds to the LO field strength of 330 mV/m. We generated our signal field with a separate SG set to 17.04341 GHz and controlled the RF output power to adjust the field strength at the cell. The LO and signal were summed using a power divider and fed into a horn antenna with a gain of 17.5 dB for this band. 

However, to avoid misrepresenting the propagation loss and effects from the glass, we calibrated the SG's using the AT splitting and Eq.~(\ref{eq:at_effect}).  These measurements cannot determine the actual field down to the field strengths of interest. However, we can determine the field at the cell location for larger RF powers and extrapolate to determine the expected field strength for the lower SG output powers, see \cite{gor3} for details. As a precaution, we also verify the linearity of the SG's down to the lower powers with a calibrated power meter and spectrum analyzer.

\begin{figure}[h!]
    \centering
    \includegraphics[width=.9\columnwidth]{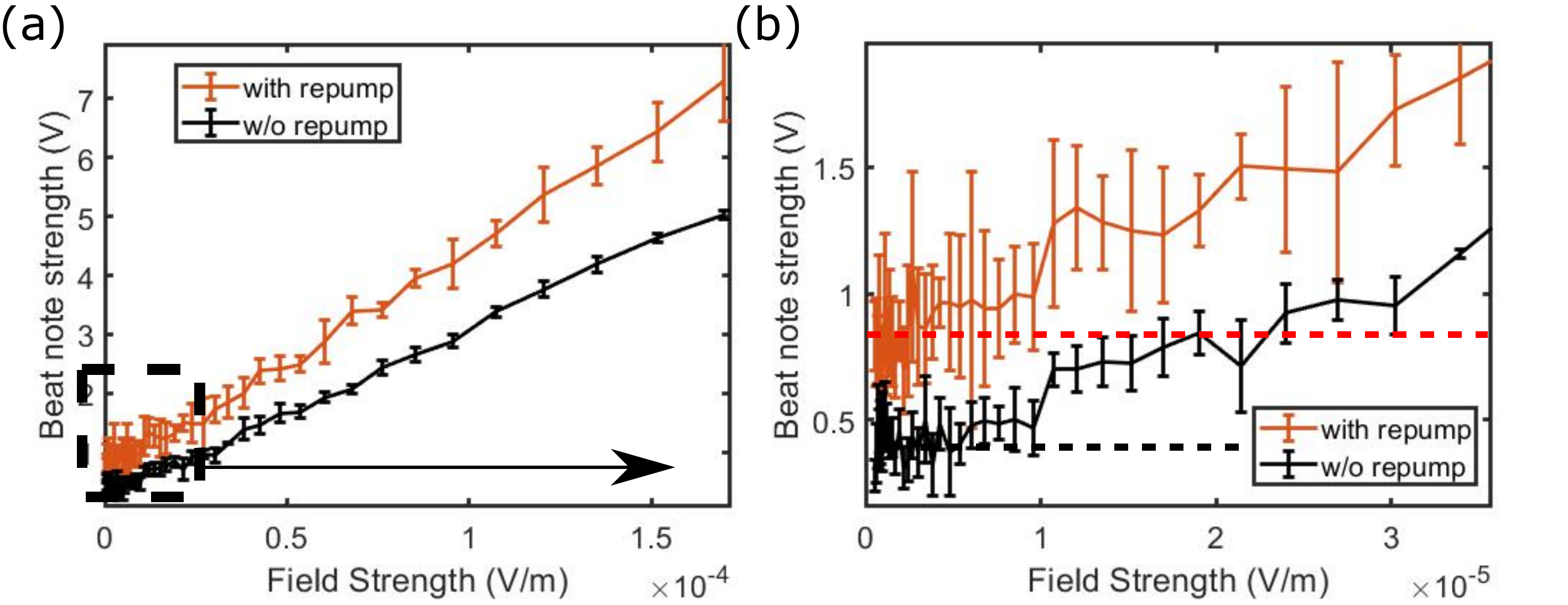}
    \caption{(a) beat-note amplitude plotted against field strength for with (red) and without (black) repump field. (b) Same plot, but zoom depicted by dashed box. The dashed lines show the respective detection limits.}
    \label{fig:cal}
\end{figure}

With this calibration, we take measurements of low output power of the SG and map the sensitivity of our sensor. We input the signal from the photodetector (Fig.~\ref{fig:beat_sample} (b)) into a lock-in amplifier which outputs the beat-note amplitude as a voltage. We use a 1 s time constant and 10 mV/A gain on the lock-in amplifier when collecting data. We took five sets of data while varying the output RF power of the signal SG (ie. signal RF field strength) for the cases with (red trace) and without (black trace) the repump field present, shown in Fig.~\ref{fig:cal} (a) and (b). The data is the average of five data sets, and the error bars are the standard deviation. The effect of the repump field is immediately apparent for large signal field strengths. The lock-in signal is 50\% larger for the trace with the repump optical field present. While we have shown that we can double the EIT peak, we limit ourselves to a probe power of 550 $\mu$W to limit the broadening effect that outweigh the effects of the gain in amplitude. The coupling laser power is 70 mW since increasing its power does not cause significant broadening.

As we apply smaller fields, we reach the detection noise floor and see that the minimum field detectable without the repump field present is $\approx$ 5 $\mathrm{\mu Vm^{-1}}$, shown by black trace in Fig.~\ref{fig:cal} (b). For the case of the data with the repump field present, the minimum field detectable is $\approx$ 3 $\mu$Vm$^{-1}$, shown by black trace in Fig.~\ref{fig:cal} (b). Since the collection time for this data is 1 s, the sensitivity for these two measurements is the same as the minimum detectable field strength. This increase in sensitivity is close to the level of gain attained on the EIT amplitude from the repump field. However, the noise floor seems to be higher when the repump is present. This noise is likely from the laser noise projection through repumping onto the probe optical field. Since the repump optical field directly alters the probe field absorption through population transfer, the fluctuations are also strongly coupled. Therefore any frequency shifts of the repump field play strongly play a role in the amplitude of the EIT. However, we are free from intensity fluctuations of the repump since the transition is saturated. 

\begin{figure}[h!]
    \centering
    \includegraphics[width=.7\columnwidth]{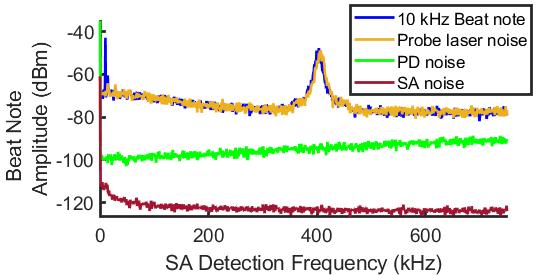}
    \caption{Noise power for atom-mixer method (blue), differential probe detection (yellow), PD dark noise (green), and SA dark noise (red). SA settings: RBW = 100 Hz, VBW = 100 Hz, and 100 averages. 
    }
    
    \label{fig:noise_comp}
\end{figure}

We utilized balanced differential detection to remove any classical noise as effectively as possible. This includes the noise introduced by the repump field. For this purpose, the repump field was split into two arms that overlapped with the signal probe and the reference probe. After analysis of the noise floor, we find that the limit in the detection is likely the probe laser noise composed of 1/f noise and photon shot noise. We compare different sources of noise measured by a spectrum analyzer in Fig.~\ref{fig:noise_comp}. The yellow trace shows the noise form the differential detection of the probe laser, in other words the shot noise limit. The detector noise (green) and SA noise (red) are several decades in power lower than the shot noise and will have no effect to the measurement noise floor. Additionally, in the presence of the blue laser and the mixer detection method (blue trace), we observe no apparent shift in the noise floor. Therefore, we can be certain the detection floor is the laser noise of the probe field. Also present in the blue trace is the expected beat-note peak at 10 kHz detection frequency.

For a more quantitative analysis, we calculate the expected noise power for the optical probe power. The photo-current variance for shot noise limited light is given in ~\cite{Fox:1001868} by
\begin{equation}
    (\Delta i)^2 = \frac{2\cdot \eta_{qe} \cdot P\cdot \lambda\cdot f_{BW}}{h\cdot c},
\end{equation}
where $h$ is Plank's constant, $c$ is speed of light, $\lambda$ is wavelength, $P$ is optical power, $f_{BW}$ is bandwidth, and $\eta_{qe}$ is the quantum efficiency of the detector. Since we are performing balanced detection with 70\% quantum efficient photo-diodes, being shot noise limited is a good assumption. The expected noise power ($P_{noise}$) on the SA in dBm is then
\begin{equation}
    P_{noise} = 10\cdot log_{10}(\frac{(G^2\cdot (\Delta i)^2}{\Omega}\cdot1000),
\end{equation}
where $G$ is PD gain, and $\Omega$ is load resistance of the spectrum analyzer. Accounting for resolution bandwidth (100 Hz), gain ($626\cdot10^3$ V/A), and resistance (100$~\Omega$), we can estimate the noise power on the spectrum analyzer. The probe power before the cell was $560~\mu W$, but actual power reaching the detector was 220 $\mu$W. The expected noise power from this is then -76 dBm. The yellow and blue traces in Fig.~\ref{fig:noise_comp} fall to this level at detection frequencies above 300 kHz. This establishes shot noise as a clear detection limit in the near future. As for the lower frequencies, we believe the dominating noise is 1/f laser noise since it is present in probe laser noise. While outside the scope of this study, there have been several studies that have demonstrated cancellation of 1/f noise down to frequencies less than 200 Hz and reach the shot noise floor~\cite{Prajapati:19,8426633}. The shot noise marks a stark quantum limit for improvement of this type of sensor. However, non-classical sources of light may provide further improvement. In ~\cite{Prajapati:21}, twin-beams were used to probe an excited atomic state. While this study explored an off-resonant effect, there are forms of squeezing that incorporated resonant enhancement.

In this manuscript we have demonstrated the enhancement of Rydberg sensing through the use of a repumping field. We showed that the repump increases the saturation threshold, and amplifies without increasing EIT linewidth. Additionally, this is done with modest repump power. By increasing the EIT amplitude, we managed to double the signal response to a signal RF field. The current limitations in this technology are from the width and height of the amplitude of the EIT. This broadening free enhancement marks progress towards better Rydberg-based technology. While we demonstrated increased sensitivity, we also identified another fundamental bound that must be surpassed. We identified the noises in the system and determined the noise floor to be the 1/f laser noise for lower frequencies and laser shot noise for higher frequencies. Future applications of quantum enhanced light will play an important role in the development of these sensors.





\section*{Acknowledgements}
We thank Dr. Svenja A. Knappe of Fieldline Inc in Boulder, CO for the use of some equipment used here. We also Dr. Eric Norrgard of NIST and Dr. Eugeniy Mikhailov of William and Mary for their useful technical discussions. This work was partially funded by the DARPA SAVaNT program and the NIST-on-a-Chip Program.

\section*{Data Availability Statement}
Data is available upon request.

\newpage
\bibliographystyle{aipnum4-1.bst}
\bibliography{atom_probe_bib}


\end{document}